# Management of localized energy in discrete nonlinear transmission lines

M.Sato and S.Yasui

Graduate School of Natural Science and Technology, Kanazawa University

Kanazawa, Ishikawa 920-1192, Japan

M. Kimura and T. Hikihara

Department of Electrical Engineering, Kyoto University

Nishikyo, Kyoto 615-8510, Japan

and

A. J. Sievers

Laboratory of Atomic and Solid State Physics, Cornell University

Ithaca, NY 14853-2501, USA

Abstract

The manipulation of locked intrinsic localized modes/discrete breathers is studied experimentally in nonlinear electric transmission line arrays. Introducing a static lattice impurity in the form of a capacitor, resistor or inductor has been used both to seed or destroy and attract or repel these localized excitations. In a nonlinear di-element array counter propagating short electrical pulses traveling in the acoustic branch are used to generate a stationary intrinsic localized mode in the optic branch at any particular lattice site. By changing the pulse polarity the same localized excitation can be eliminated demonstrating that the dynamical impurity associated with the propagating electrical pulse in the acoustic branch can trigger optical localized mode behavior.

**Introduction**

A variety of experiments on nonlinear macroscopic lattices have demonstrated[1-9] that dynamical energy localization appears because of the discreteness-nonlinearity combination[10-13]. Given the normal mode frequencies of the MEMS array, in the few 100 KHz range, a variety of time average image measurements have been reported for 1-D arrays. Studies of such intrinsic localized modes[14] (ILMs) have ranged from the breakup of a large amplitude uniform vibrational mode via a modulational instability into localized energy states; to their motion, locking to the driver, repulsion and decay[8]; to manipulation of ILMs by means of their attraction and repulsion to impurity modes produced by local laser heating[15,16]. So far such manipulation of localized dynamical energy has only been demonstrated for the micromechanical system.

Experimental observation of soliton-like propagation along nonlinear electrical transmission lines has a much longer history starting with the recurrence phenomena first observed in 1970[17] and elastic collisions between solitons observed a few years later[18]. The extensive theoretical and experimental work in this area focuses on the continuum-like soliton properties of nonlinear transmission lines[19-26]. Over the last decade two experimental studies have examined strongly localized excitations that require the discreteness of the nonlinear transmission line. Both propagating localized energy[3,27] and slowly traveling ILMs[9], induced via a modulational instability, have been reported.

Presented here is an experimental study focusing on the manipulation of ILMs in discrete nonlinear electrical transmission lines. First, a method for producing a stationary ILM by impurity seeding is demonstrated then this localized excitation is manipulated in a variety of ways in a fashion remarkably similar to that described in an earlier micromechanical study[15].

In the second part of this work a di-element transmission line is used to examine how propagating pulses in the acoustic spectrum can influence ILMs in the optic branch. Both production and destruction of ILMs are presented. It is concluded that with regard to an ILM in the optic branch the propagating electrical pulses in the acoustic branch act like dynamical lattice impurities.

**Experimental design**

      A diode is used as the nonlinear element for the transmission line shown in Fig. 1(a). Because of the small nonlinearity in a variable capacitor diode the bandwidth of the plane wave transmission line spectrum should be narrow to support well-defined ILMs. This can be accomplished either by using a large inductor as a coupling element between cells[9,22] or with a small inductor-capacitor pair used here in Fig. 1(a) [28]. Forward bias is used to maximize the nonlinearity of the diodes since the capacitance of the forward biased state (diffusion capacitance) is larger than the reverse biased one, if the diode speed (recombination lifetime of the injected carriers) is comparable to the oscillation period[29,30]. The voltage dependence of the diffusion capacitance is $C(V) = C_1 \exp(V/\phi)$, where $C_1$ is the zero voltage capacitance and $\phi$ is a constant. Integrating the stored charge $Q = Q_0 \left[ \exp(V/\phi) - 1 \right]$ with respect to voltage, where $Q_0 = C_1 \phi$, gives for lattice site n: $V_n = \phi \log(Q_n/Q_0 + 1)$. Writing the equation of motion for a unit cell in terms of charge, expanding $\ln(Q_n/Q_0 + 1)$ up to cubic order, and using the rotating wave approximation to simplify some higher order coefficients[31] gives

$$\frac{d^2}{dt^2}Q_n + \omega_1^2 Q_n - \frac{\omega_1^2}{2Q_0}Q_n^2 + \frac{\omega_1^2}{3Q_0^2}Q_n^3$$
$$+ \frac{C_2}{C_1}\left(\omega_2^2 - \omega_{driver}^2\right)\left(2Q_n - Q_{n-1} - Q_{n+1}\right)$$
$$- \frac{C_2}{C_1}\frac{1}{2Q_0}\left(\omega_2^2 - \omega_{driver}^2\right)\left(2Q_n^2 - Q_{n-1}^2 - Q_{n+1}^2\right) \qquad (1)$$
$$+ \frac{C_2}{C_1}\frac{1}{3Q_0^2}\left(\omega_2^2 - \omega_{driver}^2\right)\left(2Q_n^3 - Q_{n-1}^3 - Q_{n+1}^3\right)$$
$$= C_d \frac{d^2}{dt^2}V_d$$

where $\omega_1^2 = 1/L_1C_1$ and $\omega_2^2 = 1/L_2C_2$. The equation contains both onsite and intersite nonlinearity, with the last term the driver. Soft nonlinearity has been observed for a one-unit cell circuit demonstrating that the nonlinear component involving the quadratic charge terms are more important than the cubic terms. In addition for the parameters in Fig. 1(a) $\left|-\omega_1^2\right| > \left|-\left(C_2/C_1\right)\left(\omega_2^2 - \omega_{driver}^2\right)\right|$, i.e., the onsite quadratic term is ~ 3 larger than the intersite one at the experimental conditions.

The basic resonator unit cell with driver, illustrated in Fig. 1(a), is composed of a diode and a coil $L_1 = 300\mu H$. Among the several diodes we have tested, the variable capacitor diode (1SV100, $C_1 = 880pF$) had the largest nonlinearity. Sixteen unit cells are coupled into a ring shape via inductor $L_2 = 600\mu H$ and capacitor $C_2 = 200\,pF$. To excite the zone center mode all unit cells are driven uniformly by using the AC voltage source coupled via $C_d$. The corresponding small amplitude plane wave dispersion curve is given by

$$\omega^2 = \frac{\omega_1^2 + 2\omega_{21}^2\left(1 - \cos ka\right)}{1 + C_d/C_1 + 2C_2/C_1\left(1 - \cos ka\right)} \qquad (2)$$

where $\omega_{21}^2 = 1/L_2C_1$ with $C_d$ the small coupling capacitor ($C_d/C_1 \approx 0.04$), ignored in Eq. 1. An important function of $C_2$ is to decrease the spectral bandwidth[28]. The Eq. 2 dispersion curve is illustrated in Fig. 1(b), with the locking driver frequency represented by the dashed line.

To test the interaction between an ILM and a static defect, an impurity device and an analog switch (SW) are inserted as indicated by the box in Fig. 1(a). The switch turns the impurity on and off. The impurity is a capacitor, inductor, or resistor. The corresponding function at the particular site is to decrease the resonance frequency, increase the resonance frequency, or reduce the amplitude, respectively.

**ILM manipulation using impurities**

Figure 2 presents a number of experimental results, illustrating the possible static management pathways for such ILMs. The driving frequency and voltage are fixed at 270kHz and 2 V. The analog SW is turned on and off in a specific time interval as illustrated by the waveform shown in panel (a). The panels (b)-(h) identify the lattice site vs. time plots. An arrow denotes the impurity location for that experiment. The absolute voltage of each resonator node is represented by the gray scale (darker = larger amplitude). The results are as follows:

(b) Seeding of the ILM by a 100 pF capacitor. One ILM (dark horizontal line) remains after turning the impurity on and off. In this seeding process the oscillator keeps driving the array at a frequency slightly lower than the k=0 plane wave mode (before the ILM is generated). The low frequency impurity mode created during the ON state resonates with and is excited by the driver. When the SW is turned off, the impurity mode is replaced by the ILM, whose large amplitude supports the self-trapped excitation at that driving frequency.

(c) Attractive interaction by 45 pF capacitor. The ILM jumps to the impurity site when the SW is

turned on, and remains there after SW is turned off. The capacitance is smaller than for case (a) so an impurity mode is created closer to the bottom of the dispersion curve, and doesn't resonate to the driver. If one were to start from the no ILM state an ILM would not be seeded; however, with an ILM already present on the same side of the plane wave spectrum as the impurity mode attraction results[15].

(d) Destruction of an ILM by a resistor (100 Ω). The resistance is placed at the ILM site. It erases the ILM by reducing its amplitude.

(e) Repulsion of the ILM by an inductor (330μH). The inductance is placed at a neighboring lattice site and the then ILM hops away from that site. The impurity produces a high frequency impurity mode above the plane wave spectrum.    Since the ILM is still below the plane wave spectrum repulsion occurs[15].

(f) Destruction of the ILM by an inductor (330μH). The inductance is places at the ILM site. The nonlinearity cannot cover the large frequency difference between the high frequency impurity mode and the low frequency ILM mode.

We have checked that an ILM can be moved to any point by successive application of the attraction interaction shown in Fig. 2(c). Since the unit cell and coupling inductors are not exactly identical from cell to cell, there is a finite residual impurity effect; however, because of the transfer of the ILM to any lattice point, it can be concluded that the nonlinear effect associated with maintaining the ILM at any lattice site is larger than any residual lattice inhomogeneity.

The seeding, attraction, and repulsion effects shown in Fig. 2(b), (c) and (e) have already been demonstrated in MEMS simulations, and by experiments[16]. Destruction of an ILM by a resistor or an inductor [shown in Fig. 2(d) and (f)] are new phenomena. Controlling an

ILM in a nonlinear electric transmission lines has more options than with the MEMS cantilever array, where only a low frequency impurity mode could be produced.

**Control of ILMs with acoustic-like pulses**

A DC bias voltage can be applied to the diodes by inserting another capacitance $C_3 = 0.022 \mu F$ as shown in Fig. 3(a) in series with $L_1$. This produces a new voltage node between $L_1$ and $C_3$, in addition to the original node at the diode, so the number of degree of freedom in the unit cell is now two. This circuit can be separated into two circuits as shown in and can be used to generate ILMs. The circuit for low frequencies (lower part) can be used for acoustic pulse propagation. The resulting two branches in the dispersion map are shown in Fig.

Fig. 3(b). At low frequencies, the impedances of $C_2$ and $C_3$ are large, and can be ignored. At high frequencies, the impedance of $C_3$ is small, and can be treated as short. The circuit for high frequencies shown in the upper part of this figure is the same as for the circuit shown in Fig. 1(a),

3(c). (The low-frequency circuit is similar to the one used to examine discreteness effects on the propagation of nonlinear waves[22,27]). With the addition of a positive DC bias voltage, either positive or negative pulses can be transmitted. To reduce pulse reflections, a terminating resistance R is inserted, as shown in Fig. 3(a). For the experiments described here a second pulse generator and resistor are connected at the opposite end.

With this experimental arrangement the interaction between ILMs in the upper branch and pulses propagating in the lower branch have been studied. When a single pulse of sufficient negative amplitude is sent through the transmission line a number of locked ILMs can be generated. A corresponding positive pulse can eliminate these same ILMs. Of interest here is the effect produced by colliding pulses when neither one by itself has sufficient amplitude to produce an ILM.

Figure 4 shows experimental results for ILMs associated with colliding negative or positive pulses injected from both edges of the lattice. Such pulses collide at some point in the lattice, and this intersection point can be adjusted by changing one injection time relative to the other. Figures 4(a) and (b) show the generation of a locked ILM at different lattice sites by two negative pulses, while Figs. 4(c) and 4(d) illustrate that ILM destruction can be produced by positive pulses when they intersect at the ILM lattice site. In both kinds of experiments, an individual pulse height isn't large enough by itself to produce the observed effect.

To understand these processes note that the acoustic pulse voltage changes slowly as a function of time compared to the ILM; hence, it modulates the capacitance locally, as would a static impurity mode. As an example, when the DC voltage of a cell is changed from 0 to 1 to 2V, $C_1$ decreases from 880, to 550 to 370 pF, respectively. For a 1 V bias with $C_1 = 550\,pF$ the regular plane wave band would extend from 380 to 430 kHz. In this case an 880 pF defect would make a localized gap mode at 340 kHz (a 370 pF defect, a localized mode at 460 kHz). Thus a propagating impurity mode moves with the electric pulse so that negative and positive voltage pulses form low and high frequency running impurity modes, respectively. The ILM appears when two counter-propagating negative pulses coexist in space, with the frequency of the concomitant impurity mode close enough to the ILM locking frequency to generate the ILM. Conversely, with positive voltage pulses the high frequency of the impurity mode is far removed from the locking frequency and thus destroys the ILM. Such ON/OFF ILM effects are similar to those demonstrated in Fig. 2 but now ILM management is achieved externally. Recently an interaction between acoustic plane wave and ILMs has been studied via simulations[32].

**Summary**

We have shown that ILMs can be generated and manipulated by local impurity switching in nonlinear discrete electrical transmission lines. Managing ILMs from outside of the lattice with acoustic pulses has also been experimentally demonstrated for the first time. Such manipulation of an electrical ILM in these nonlinear transmission lines indicates that similar properties may be expected for a monolithic integrated circuit[20], a sonic Helmholtz array[33] or a line of nanoelectronic resonators[34]. Since there is no fundamental difference between voltage waves in our nonlinear networks and electromagnetic waves in nonlinear layered structures[35,36] or in nonlinear photonic crystals[37,38] there should be many theoretical/experimental analogues to the processing of such localized energy.

**Acknowledgment**

This work was supported in part by NSF DMR 0301035, by DOE DE-FG02-04ER46154 and by JSPS-Grant-in-Aid for Scientific Research (B)18340086.

# References

[1]     W.-Z. Chen, Phys. Rev. B **49**, 15063 (1994).

[2]     S. Lou and G. Huang, Mod. Phys. Lett. B **9**, 1231 (1995).

[3]     P. Marquié, J. M. Bilbault, and M. Remoissenet, Phys. Rev. E **51**, 6127 (1995).

[4]     P. Binder, D. Abraimov, A. V. Ustinov, S. Flach, and Y. Zolotaryuk, Phys. Rev. Lett. **84**, 745 (2000).

[5]     E. Trías, J. J. Mazo, and T. P. Orlando, Phys. Rev. Lett. **84**, 741 (2000).

[6]     J. W. Fleischer, T. Carmon, M. Segev, N. K. Efremidis, and D. N. Christodoulides, Phys. Rev. Lett. **90**, 023902 (2003).

[7]     D. Mandelik, H. S. Eisenberg, Y. Silberberg, R. Morandotti, and J. S. Aitchison, Phys. Rev. Lett. **90**, 253902 (2003).

[8]     M. Sato, B. E. Hubbard, A. J. Sievers, B. Ilic, D. A. Czaplewski, and H. G. Craighead, Phys. Rev. Lett. **90**, 044102 (2003).

[9]     R. Stearrett and L. Q. English, arXiv **0706**, 1212 (2007).

[10]    A. J. Sievers and J. B. Page, in *Dynamical Properties of Solids: Phonon Physics The Cutting Edge*, edited by G.K. Norton and A.A. Maradudin (North Holland, Amsterdam, 1995), Vol. VII, p. 137.

[11]    S. Flach and C. R. Willis, Phys. Repts. **295**, 182 (1998).

[12]    R. Lai and A. J. Sievers, Phys. Repts. **314**, 147 (1999).

[13]    D. K. Campbell, S. Flach, and Y. S. Kivshar, Physics Today **57**, 43 (2004).

[14]    A. J. Sievers and S. Takeno, Phys. Rev. Lett. **61**, 970 (1988).

[15]    M. Sato, B. E. Hubbard, A. J. Sievers, B. Ilic, and H. G. Craighead, Europhys. Lett. **66**, 318 (2004).

[16]    M. Sato, B. E. Hubbard, and A. J. Sievers, Rev. Mod. Phys. **78**, 137 (2006).

[17]    R. Hirota and K. Suzuki, J. Phs. Soc. Jpn. **28**, 1366 (1970).

[18]    R. Hirota and K. Suzuki, Proc. IEEE **61**, 1483 (1973).

[19]    A. C. Scott, *Active Nonlinear Wave Propagation in Electronics* (Wiley-Interscience, New York, 1970).

[20]    D. Jäger, Int. J. Electronics **58**, 649 (1985).

[21]    M. A. Malomed, Phys. Rev. A **45**, 4097 (1992).

[22]    P. Marquié, J. M. Bilbault, and M. Remoissenet, Phys. Rev. E **49**, 828 (1994).

[23]    T. Kuusela, Chaos, Solitons & Fractals **5**, 2419 (1995).

[24]    A. Scott, *Nonlinear Science: Emergence and Dynamics of Coherent Structures* (Oxford University Press, New York, 1999).

[25]    B. Z. Essimbi and D. Jäger, J. Phys. D: Appl. Phys. **39**, 390 (2006).

[26]    E. Afshari, H. S. Bhat, A. Hajimiri, and J. E. Marsden, J. Appl. Phys. **99**, 054901 (2006).

[27]    M. Remoissenet, *Waves Called Solitons*, Third ed. (Springer-Verlag, Berlin, 1999), p. 76.

[28]    K. Fukushima, M. Wadati, and Y. Narahara, J. Phys. Soc. Jpn. **49**, 1593 (1980).

[29]    S. Wang, *Solid State Electronics* (McGraw-Hill, New York, 1966), Vol. Ch. 6.

[30]    R. Van Buskirk and C. D. Jeffries, Phys. Rev. A **31**, 3323 (1985).

[31]    S. R. Bickham, S. A. Kiselev, and A. J. Sievers, Phys. Rev. B **47**, 14206 (1993).

[32]    Y. Doi and A. Nakatani, Key Eng. Mat. **340-341**, 997 (2007).

[33]    X. Hu, C. T. Chan, and J. Zi, Phys. Rev. E **71**, 055601(R) (2005).

[34]    H. G. Zhang, Z. Chen, L. T. X, F. J. Wang, and K. Saito, J. Electrochem. Soc. **154**, H124 (2007).

[35]    W. Chen and D. L. Mills, Phys. Rev. Lett. **58**, 160 (1987).

[36]    D. L. Mills, *Nonlinear Optics: Basic Concepts* (Springer-Verlag, Berlin, 1991), p. 139.

[37]    A. R. McGurn, Phys. Letters A **251**, 322 (1999).

[38]    A. R. McGurn, Phys. Rev. B **65**, 1 (2002).

**Figure captions**

Fig. 1 (a) Electrical transmission line for controlling ILMs by impurity switching.

Diode(1SV100) is used for the nonlinear element. Each unit cell is composed of a diode and inductor $L_1$ for the resonating elements, and $L_2$ and $C_2$ for the coupling elements. In the circular device the unit cells are uniformly driven by an AC oscillator via the small coupling capacitor $C_d$. To introduce a defect into the lattice, an impurity device, indicated by the rectangle, is switched ON and OFF at a particular site.

(b) Small amplitude plane wave dispersion curve of the circuit shown in panel (a). The dotted

line identifies the driver frequency of the locked ILM.

Fig. 2. Demonstration of ILM management by the introduction of different defects. Control wave

of the SW is shown in panel (a). In panels (b)-(f) the amplitudes are represented by the gray scale.

The impurity location in the lattice is represented by arrows on the left side. (b) Seeding of the

ILM by 100 pF capacitor. (c) Attraction by a 45 pF capacitor. (d) Destruction by $100\,\Omega$ resistor.

(e) Repulsion by 330 $\mu H$ inductor. (f) Destruction by 330 $\mu H$ inductor. The amplitude of the

ILM is about 4.2 V, while the background in the rest of the array is 2.3 V.

Fig. 3(a) Circuit diagram for acoustic pulse control of optical ILMs. A second degree of freedom

for the unit cell is provided by adding $C_3$.

(b) Equivalent circuit for the pulse control experiment. Upper network is for the optical ILMs at higher frequencies, and the lower network is for the acoustic pulse at lower frequencies. Diodes are indicated as variable capacitors. At each point, the capacitance is shared by two networks. The upper branch is almost the same as the branch shown in Fig. 1(a), and can be used for generating ILMs. The presence of $C_3$ permits a DC bias (0.5~1V) and the possibility of either positive or negative pulse propagation. A terminating resistor R and pulse generator is attached to each end of the array.

(c) The two branches of the plane wave dispersion curve of the circuit shown in panel (a). Dotted line identifies the driver frequency.

Fig. 4.    Optical ILM control by acoustic pulses. The AC driver at the ILM frequency is kept ON. Voltage amplitudes relative to the DC bias are displayed.    Panels (a) and (b) show ILM generation at the intersection of negative pulses which are injected at the edges. The ILM generation lattice site is controlled by the relative delay time between pulses. Dark slanting lines are traces of the negative pulses, horizontal line, the ILM. Panels (c) and (d) show the destruction effect of two positive colliding pulses. In panel (c), the pulse intersection is not at the ILM lattice site and it survives. Panel (d) shows that when two positive pulses collide at the ILM position, it is destroyed. Experimental input conditions: 1.2V amplitude for the 324 kHz ILM driver, 1V DC

bias, -2V, 3.5 $\mu s$ for negative pulses, and +2.6V, 2 $\mu s$ for positive pulses. The transmitting pulse widths of the system shown in the figure are about 15 $\mu s$.

Figures

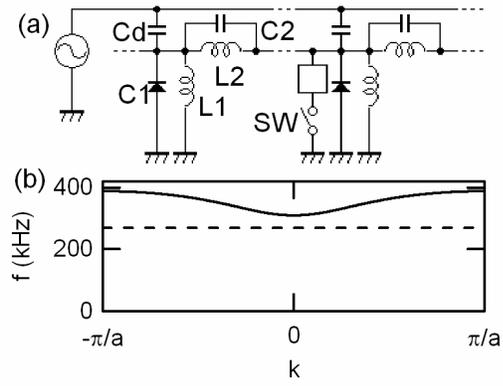

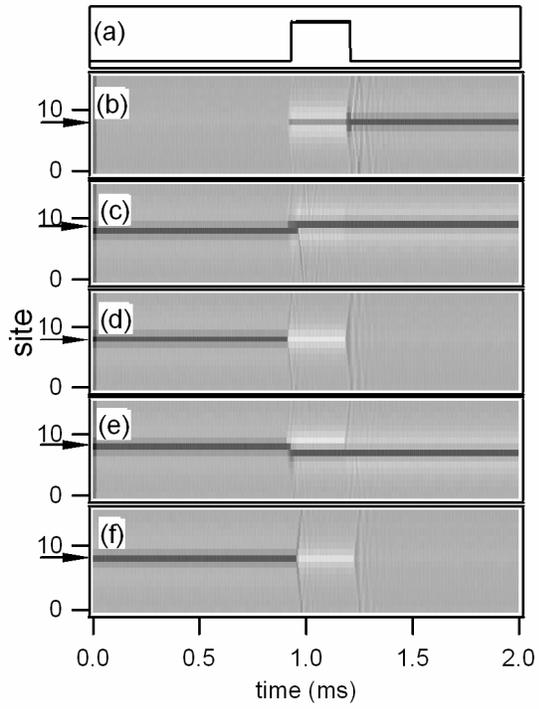

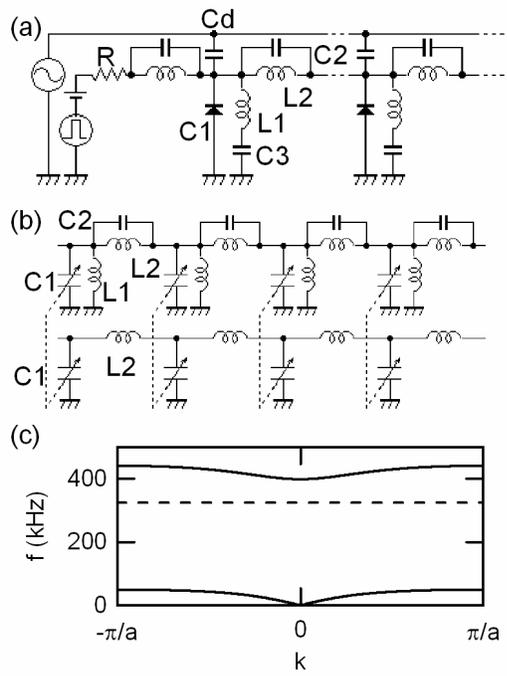

(a)

(b)

(c)

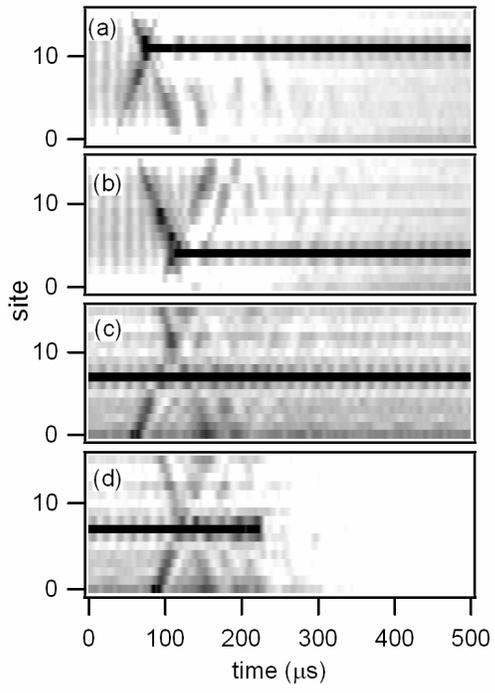